\documentclass[twocolumn]{aastex61}
\usepackage{float,amsmath,multirow,mathtools}
\usepackage{color}
\usepackage[version=3]{mhchem}
\newcommand{\isoc}{$^{13}$C}

\begin{document}

\title{Detection of HC$_5$N and HC$_7$N Isotopologues in TMC-1 with the Green Bank Telescope}
\author{Andrew M. Burkhardt}
\affiliation{Department of Astronomy, University of Virginia, Charlottesville, VA 22903, USA}
\author{Eric Herbst}
\affiliation{Department of Chemistry, University of Virginia, Charlottesville, VA 22903, USA}
\affiliation{Department of Astronomy, University of Virginia, Charlottesville, VA 22903, USA}
\author{Sergei Kalenskii}
\affiliation{Astro Space Center, Lebedev Physical Institute, Russian Academy of Sciences, Moscow, Russia}
\author{Michael C. McCarthy}
\affiliation{Harvard-Smithsonian Center for Astrophysics, Cambridge, MA 02138, USA}
\author{Anthony J. Remijan}
\affiliation{National Radio Astronomy Observatory, Charlottesville, VA 22903, USA}
\author{Brett A. McGuire}
\altaffiliation{B.A.M. is a Hubble Fellow of the National Radio Astronomy Observatory}
\affiliation{National Radio Astronomy Observatory, Charlottesville, VA 22903, USA}
\affiliation{Harvard-Smithsonian Center for Astrophysics, Cambridge, MA 02138, USA}
\correspondingauthor{Andrew M. Burkhardt \& Brett A. McGuire}
\email{amb3au@virginia.edu \& bmcguire@nrao.edu}

\begin{abstract}
We report the first interstellar detection of DC$_7$N and six $^{13}$C-bearing isotopologues of HC$_7$N toward the dark cloud TMC-1 through observations with the Green Bank Telescope, and confirm the recent detection of HC$_5$$^{15}$N. For the average of the $^{13}$C isotopomers, DC$_7$N, and HC$_5$$^{15}$N, we derive column densities of 1.9(2)$\times$10$^{11}$, 2.5(9)$\times$10$^{11}$, and 1.5(4)$\times$10$^{11}$ cm$^{-2}$, respectively. The resulting isotopic ratios are consistent with previous values derived from similar species in the source, and we discuss the implications for the formation chemistry of the observed cyanopolyynes. Within our uncertainties, no significant $^{13}$C isotopomer variation is found for HC$_7$N, limiting the significance CN could have in its production. The results further show that, for all observed isotopes, HC$_5$N may be isotopically depleted relative to HC$_3$N and HC$_7$N, suggesting that reactions starting from smaller cyanopolyynes may not be efficient to form HC$_{n}$N. This leads to the conclusion that the dominant production route may be the reaction between hydrocarbon ions and nitrogen atoms. 
\end{abstract}




\section{Introduction}\label{sec:intro}
Carbon-chain molecules are a critically important family within the interstellar medium (ISM); they represent $\sim$40\% of all detected species and play a major role in the formation of more complex chemistry. Carbon chains with interstellar detections include: carbenes \citep{McCarthy:1997uv}, polyynes \citep{Irvine:1981, Bell:1997vj, Snyder:2006, Remijan:2006tw}, unsaturated hydrocarbons \citep{Cernicharo:2001}, and the newly detected HC$_n$O family \citep{McGuire:2017}. Furthermore, it has been suggested that they may be important precursors to the formation of polycyclic aromatic hydrocarbons (PAHs) \citep{Guzman-Ramirez:2011}, which are likely to be routine targets of observation when the James Webb Space Telescope is launched in 2019 \citep{Kirkpatrick:2017}. 

Reactions of unsaturated carbon-chain molecules (i.e. species whose available carbon valence electrons are not all bonded to atoms so that the chain contains more double and triple carbon-carbon bonds) are often efficient in the ISM, but have many product channels for which branching ratios are not known. Because of the lack of available laboratory measurements of the dominant reaction mechanisms and rate coefficients for these branching fractions, the ability to directly investigate these pathways through interstellar observations is therefore appealing. One way to probe the underlying chemistry is through the study of the isotopologues (molecules that differ only in isotopic composition (e.g. HCCCN vs H$^{13}$CCCN)) and isotopomers (molecules that contain the same isotopic composition, but differ in the isotope positions (e.g.  H$^{13}$CCCN vs  HCC$^{13}$CN)) of a species. For many molecules, specific relative isotopic fractions or isotopomer configurations can possibly constrain the dominant production method or precursors. 

TMC-1, one of the prototypical dark cloud cold cores, has been the subject of intense astrochemical study. Observationally, it has been the source of many new molecular detections, including a large fraction of the known unsaturated carbon-chains \citep{Kaifu:2004tk,Snyder:2006,McCarthy:2006tk,Remijan:2006tw,McGuire:2017}. Furthermore, TMC-1 is dynamically stable, characterized by narrow line widths ($\sim$0.3 km s$^{-1}$), cold excitation temperatures (5-10 K), and a low line density ($\sim$1 line per 200 km s$^{-1}$ for reasonable integration times with the GBT), making it ideal for the unambiguous detection of new molecules. Finally, because of its simple physical history, it is an ideal source to test chemical network models \citep{Hasegawa:1992,Herbst&Millar:2008,Ruaud:2016,Majumdar:2017}. 

Cyanopolyynes, a family of linear molecules of the form HC$_n$N (where $n$\,=\,3, 5, 7, etc., henceforth) with alternating single and triple-bonded carbon atoms, have been detected in cold dark clouds \citep{Broten:1978}, the expanding envelopes of evolved stars \citep{Bell:1992}, and even external galaxies \citep{Mauersberger:1990}. It has been shown that, unlike many carbon-chains, the $^{12}$C/$^{13}$C ratio for HC$_5$N is constant even into subsequent stages of star formation through observations of warm carbon-chain chemistry in the low-mass star-forming region L1527 \citep{Araki:2016, Taniguchi:2016b}. This finding implies that the formation of cyanopolyynes may occur primarily under dark cloud conditions. These species then remain as relics in subsequent stages of star formation, and have a unique underlying chemistry compared with other carbon-chain molecules. However, recent observations of a more evolved core, L134N, suggest that other formation pathways may dominate at later times \citep{Taniguchi:2017b}. 

Of particular interest, \citet{Loomis:2016js} recently discussed the non-detection of HC$_{11}$N, which deviates from the log-linear abundance vs molecular size trend seen for smaller cyanopolyynes \citep{Bujarrabal:1981, Bell:1997vj, Ohishi&Kaifu:1998, Remijan:2006tw}. Although this trend was previously thought to arise from a consistent set of gas-phase reactions that add carbons directly to smaller HC$_n$N species \citep{Remijan:2005, Winnewisser&Walmsley:1979, Bujarrabal:1981, Fukuzawa:1998}, it was proposed by \citet{Loomis:2016js} that cyclisation processes may need to be considered to accurately explain this deviation. It is clear, therefore, that the chemistry of this family of species is not fully understood, especially at larger molecular sizes, and further study is needed. 

For the cyanopolyyne family, \citet{Takano:1998} and \citet{Taniguchi:2016} discussed the potential prominence of three formation routes, among others, for a given molecule (HC$_n$N), which could each result in different $^{12}$C/$^{13}$C fractionations. Each of these numbered mechanisms is discussed below, and, for clarity, sources of carbon atoms which could result in $^{13}$C fractionation are traced from reactants to products in example reactions in \textcolor{red}{red}.\\

\noindent \emph{\underline{Mechanism 1} - The reaction between hydrocarbon molecules and the CN radical, including}
\begin{equation}
 \text{C}_{n-1}\text{H}_2 + \textbf{\textcolor{red}{C}}\text{N} \rightarrow \text{HC}_{n-1}\textbf{\textcolor{red}{C}}\text{N} + \text{H} \label{eq:cn}. 
\end{equation}
Here, the difference between the isotopic fractions of the CN carbon atom and the carbene C$_n$H$_2$ carbon atoms results in asymmetric fractionation along the chain \citep{Herbst&Leung:1990, Fukuzawa:1998}.

\noindent \emph{\underline{Mechanism 2} - Reactions of the next-smallest cyanopolyyne (e.g. HC$_5$N vs HC$_3$N) with hydrocarbons, such as}
\begin{equation}
\text{C}_2\text{H} + \text{H}\textbf{\textcolor{red}{C$_{n-2}$}}\text{N} \rightarrow \text{HC}_2\textbf{\textcolor{red}{C$_{n-2}$}}\text{N} + \text{H}.  \label{eq:hcn-2n}
\end{equation}
For this case, many of the isotopomers would have similar $^{12}$C/$^{13}$C ratios to their corresponding isotopomer of the precursor cyanopolyyne, with potentially some small variations depending on the precursor hydrocarbon \citep{Schiff&Bohme:1979, Huntress:1977}.

\noindent \emph{\underline{Mechanism 3} - Reactions of nitrogen atoms and hydrocarbon ions containing the same number of carbon atoms.} One such example is
\begin{alignat}{4}
&\textbf{\textcolor{red}{C$_n$}}\text{H}_{3}^{+} &&+ \text{N}  &&\rightarrow \text{H}_2\textbf{\textcolor{red}{C$_n$}}\text{N}^+ &&+ \text{H} \nonumber \\
&\text{H}_2\textbf{\textcolor{red}{C$_n$}}\text{N}^+ &&+ e^{-}&&\rightarrow \text{H}\textbf{\textcolor{red}{C$_n$}}\text{N} &&+ \text{H} \label{eq:n}.
\end{alignat}
In this scenario, $^{12}$C/$^{13}$C fractionation would be set by this precursor ion. Assuming the ion's carbon atoms are sufficiently scrambled, this manifests as no significant variations among the isopotomers' ratios \citep{Herbst:1983, Herbst:1984,Knight:1986}.

Previously, the $^{12}$C/$^{13}$C fractionation had only been studied for cyanoacetylene (HC$_3$N) and cyanodiacetylene (HC$_5$N). For HC$_3$N, \citet{Takano:1998} found that toward TMC-1 there was a $\sim$40\% abundance enhancement of the isotopomer with the $^{13}$C residing next to the nitrogen atom (HC$_2$$^{13}$CN) relative to the other two species, suggesting that the primary formation route could be the neutral reactions between the abundant CN and C$_2$H$_2$ (Mechanism \ref{eq:cn}). This enhancement in the $^{13}$CN is thought to result from the exothermic exchange reaction given by
\begin{equation}
	\text{CN} + \,^{13}\text{C}^+ \rightarrow \,^{13}\text{CN} + \text{C}^+ + \Delta E(34 K) \label{eq:13cn}.
\end{equation}  
Because the reaction is exothermic, only the forward process is efficient at the cold temperatures within dark clouds, which results in an enhanced $^{13}$C fractionation in CN compared with the carbene precursors whose exchange reactions are much less efficient \citep{Benson:1989, Watson:1976}. This enhancement of $^{13}$C in CN relative to measured solar system isotopic fractions has been ubiquitously observed in Galactic molecular clouds \citep{Milam:2005}.

Meanwhile, \citet{Taniguchi:2016} found that there was no significant difference in the abundance of the \isoc-isotopomers of HC$_5$N toward TMC-1, suggesting that the primary formation route for HC$_5$N could be from reactions of N with hydrocarbon ions (such as C$_5$H$_3^+$, C$_5$H$_4^+$, and C$_5$H$_5^+$). 

It is also important to compare these results to the dominant formation routes within chemical network models. Significant work has been done for this molecular family by \citet{Loomis:2016js} and \citet{McGuire:2017} who adapted the \textsc{kida} network within \textsc{nautilus} \citep{Ruaud:2016}. At the model's time of best agreement, multiple formation routes significantly contributed (>30\%) to the formation of cyanopolyynes, many of which do not necessarily agree with the observational constraints.

We have recently performed deep observations of TMC-1, which has resulted in the interstellar detection of several new molecules \citep{McGuire:2017}. Here, we present the detection of six of the seven possible $^{13}$C-bearing isotopomers of HC$_7$N, as well as DC$_7$N. In addition, we confirm the recent detection of HC$_5$$^{15}$N. The observations are presented in \S\ref{sec:observations}, a review of the laboratory spectroscopy is given in \S\ref{sec:spectroscopy}, the results and analysis are discussed in \S\ref{sec:results}, and a discussion of the astrochemical implications is given in \S\ref{sec:discussion}.

\section{Observations}\label{sec:observations}
Observations, described previously by \citet{McGuire:2017}, toward TMC-1 were performed on the 100 m Robert C. Byrd Green Bank Telescope (GBT) in Green Bank, WV with the K-band Focal Plane Array (KFPA) along with the Versatile GBT Astronomical Spectrometer (VEGAS) spectrometer backend. 
The beam size varied from 32-40\arcsec across the observed frequency range, with a beam efficiency of $\sim$\added{0.92}.
The VEGAS backend was configured for 187.5 MHz bandwidth and 1.4 kHz (0.02 km s$^{-1}$) spectral resolution. In two separate frequency setups, a total of ten individual passbands were observed for a total of 1875 MHz of spectral coverage between 18 and 24 GHz. The observations were centred on $\alpha$(J2000) = 04$^{\rm{h}}$41$^{\rm{m}}$42\fs 5, $\delta$(J2000) = 25\degr 41\arcmin 27\farcs 0, with pointing corrections performed hourly with an estimated uncertainty of $\sim$2\arcsec. The system temperatures ranged between 40-80 K during the observations.

Position-switching mode was used with a 120 s ON-OFF cadence and a position 1\degr offset from the target. In total, each of the ten frequency windows were observed between $\sim$7.5 and 15 hours on source. Data reduction was performed using the \textsc{GBTIDL} package.  The data were placed on the atmosphere-corrected $T_A^*$ scale \citep{Ulich&Haas:1976} and averaged.  A polynomial fit was used to remove the baseline. Subsequent smoothing to a spectral resolution of 5.7 kHz ($\sim$0.08 km s$^{-1}$) improved the signal to noise ratio (SNR) in the weaker features while maintaining at least 3 channels sampling across the narrowest spectral feature observed. This resulted in final RMS noises of 3-5 mK (Table \ref{tab:transitions})

\section{Spectroscopy}
\label{sec:spectroscopy}
For the new species detected here, \citet{McCarthy:2000eo} measured the pure rotational spectra of the isotopologues of several cyanopolyynes, including HC$_7$N, between 6 and 17 GHz, and resolved the nitrogen hyperfine splitting. The rotational spectrum for HC$_5$$^{15}$N was measured by \citet{Bizzocchi:2004}. The corresponding quantum transitions, frequencies (MHz), line strengths (D$^2$), and upper-level energies (K) for transitions falling within our observational coverage are shown in Table \ref{tab:transitions}. 

\begin{table*}
\centering
\scriptsize
\caption{Measured and observed frequencies of detected HC$_7$N and HC$_5$N isotopomer and isotopologue transitions covered in this work and pertinent line parameters from Gaussian fits.}
\begin{tabular}{c c c c c c c c}
\hline\hline
Species				&	$J^{\prime} \rightarrow J^{\prime\prime}$	&	Frequency$^{a}$	&	$V_{\text{lsr}}\,^{b}$	&	$\Delta T_A^*$$^c$	&	$\Delta V$	&	$S_{ij}\mu^2$	&	$E_{u}$	\\
					&									&	(MHz)			&	(km s$^{-1}$)		&	(mK)				&	(km~s$^{-1}$)	&	(Debye$^2$)	&	(K)		\\
\hline
HC$_7$N				&	17$\rightarrow$16					&	19175.959		&	5.81				&	909(3)			&	0.474(2)			&	394.9		&	8.283   \\
					&	18$\rightarrow$17					&	20303.946		&	5.83				&	978(2)			&	0.458(1)			&	418.2		&	9.257   \\
\\
HC$_6$$^{13}$CN		&	18$\rightarrow$17					&	20071.326		&	5.84				&	12(1)			&	0.36(5)			&	418.2	&	9.151   \\
					&	19$\rightarrow$18	&	21186.389		&	5.86	&	13.2(9)	&	0.46(3)	&	441.4	&	10.168   \\
					&	20$\rightarrow$19	&	22301.449	&	5.84	&	19(3) 	&	0.31(6)	&	464.7	&	11.238   \\\\
HC$_5$$^{13}$CCN	&	17$\rightarrow$16		&	19102.044	&	5.84	&	16(1)		&	0.52(5)	&	394.9	&	8.251   \\
\\
HC$_4$$^{13}$CC$_2$N	&	18$\rightarrow$17	&	20294.271	&	5.89	&	12(1)		&	0.54(8)	&	418.3	&	9.253   \\
\\
HC$_3$$^{13}$CC$_3$N	&	17$\rightarrow$16	&	19165.136	&	5.70	&	9.9(4)	&	0.50(2)	&	394.9	&	8.278   \\
\\
HC$_2$$^{13}$CC$_4$N	&	17$\rightarrow$16	&	19097.498	&	5.81	&	12.8(3)	&	0.48(1)	&	394.9	&	8.249   \\
					&	18$\rightarrow$17	&	20220.870	&	5.76	&	14(3)		&	0.32(8)	&	418.2	&	9.219   \\
\\
HC$^{13}$CC$_5$N		&	18$\rightarrow$17	&	20063.864	&	5.80	&	9(1)		&	0.51(7)	&	418.1	&	9.148   \\
					&	20$\rightarrow$19	&	22293.157	&	5.76	&	18(3)		&	0.42(8)	&	464.6	&	11.234   \\
\\
H$^{13}$CC$_6$N		&	21$\rightarrow$20	&	23168.899	&	-$^{d}$		&	$<$14.8$^{d}$		&	0.4$^{d}$	&	487.4		&	12.231   \\
\\
DC$_7$N	&	19$\rightarrow$18	&	20721.873	&	5.97	&	15(3)	&	0.34(7)	&	441.4	&	9.945   \\
		&	20$\rightarrow$19	&	21812.486	&	5.84	&	12.2(6)	&	0.62(4)	&	464.7	&	10.992   \\
		&	21$\rightarrow$20	&	22903.097	&	5.92	&	16(1)	&	0.49(6)	&	487.9	&	12.091   \\
\hline
HC$_5$N	&	8$\rightarrow$7,\,$F$=8$\rightarrow$8	&	21299.750	&	5.82	&	32(3)	&	0.43(4)	&	0.781	&	4.600   \\
\\
		&	8$\rightarrow$7,\,$F$=7$\rightarrow$6	&	21301.245	&		&		&		&	43.3	&	4.600   \\
		&	8$\rightarrow$7,\,$F$=8$\rightarrow$7	&	21301.261	&	5.82	&	2489(11)	&	0.650(3)	&	49.2	&	4.600   \\
		&	8$\rightarrow$7,\,$F$=9$\rightarrow$8	&	21301.272	&		&		&		&	55.9	&	4.600   \\
\\
		&	8$\rightarrow$7,\,$F$=7$\rightarrow$7	&	21302.970	&	5.85	&	25(2)	&	0.54(6)	&	0.781	&	4.600   \\
\\
HC$_5$$^{15}$N	&	8$\rightarrow$7	&	20778.180	&	5.78	&	16.9(9)	&	0.46(3)	&	150.0	&	4.487   \\
\\
HC$_4$$^{13}$CN	&	7$\rightarrow$6	&	18454.489	&	5.80	&	28(1)	&	0.76(4)	&	131.2	&	3.543   \\
HC$_3$$^{13}$CCN	&	8$\rightarrow$7	&	21281.792	&	5.82	&	40(2)	&	0.59(3)	&	151.4	&	4.596   \\
HC$_2$$^{13}$CC$_2$N	&	8$\rightarrow$7	&	21279.200	&	5.81	&	42(2)	&	0.67(4)	&	150.0	&	4.596   \\
HC$^{13}$CC$_3$N	&	7$\rightarrow$6	&	18447.612	&	5.81	&	20(1)	&	0.79(5)	&	130.0	&	3.541   \\
H$^{13}$CC$_4$N	&	8$\rightarrow$7	&	20746.761	&	5.81	&	42(1)	&	0.59(2)	&	150.0	&	4.481   \\
\\
DC$_5$N	&	8$\rightarrow$7	&	20336.870	&	5.82	&	52(2)	&	0.47(2)	&	150.001	&	4.392   \\
		&	9$\rightarrow$8	&	22878.963	&	5.86	&	59(2)	&	0.54(2)	&	168.729	&	5.490   \\
\hline
\hline
\multicolumn{8}{l}{$^{a}$\citet{McCarthy:2000eo} had a 1$\sigma$ experimental uncertainty of $\sim$2 kHz.}\\
\multicolumn{8}{l}{~~  \citet{Bizzocchi:2004} had a 1$\sigma$ experimental uncertainty of $\sim$15 kHz.}\\
\multicolumn{8}{l}{$^{b}$1$\sigma$ uncertainties from Gaussian fits are $\sim$0.5 kHz (0.08 km s$^{-1}$).  Given the SNR of the detected lines ($\sim$3-5)}\\
\multicolumn{8}{l}{~~and the linewidth, we estimate the uncertainty in the observed line centres to be $\sim$3.7 kHz.}\\
\multicolumn{8}{l}{$^{c}$1$\sigma$ uncertainty of the Gaussian fit to each line given.}\\
\multicolumn{8}{l}{$^{d}$Upper limit of line peak set by 3$\times$RMS at transition frequency.}\\
\multicolumn{8}{l}{\,\,~For purposes of $N_T$ calculations, $\Delta V$ was estimated to be 0.4 km s$^{-1}$.}
\end{tabular}
\label{tab:transitions}
\end{table*}

\section{Results and Analysis}\label{sec:results}
\begin{figure}
\centering
\includegraphics[width=0.5\textwidth]{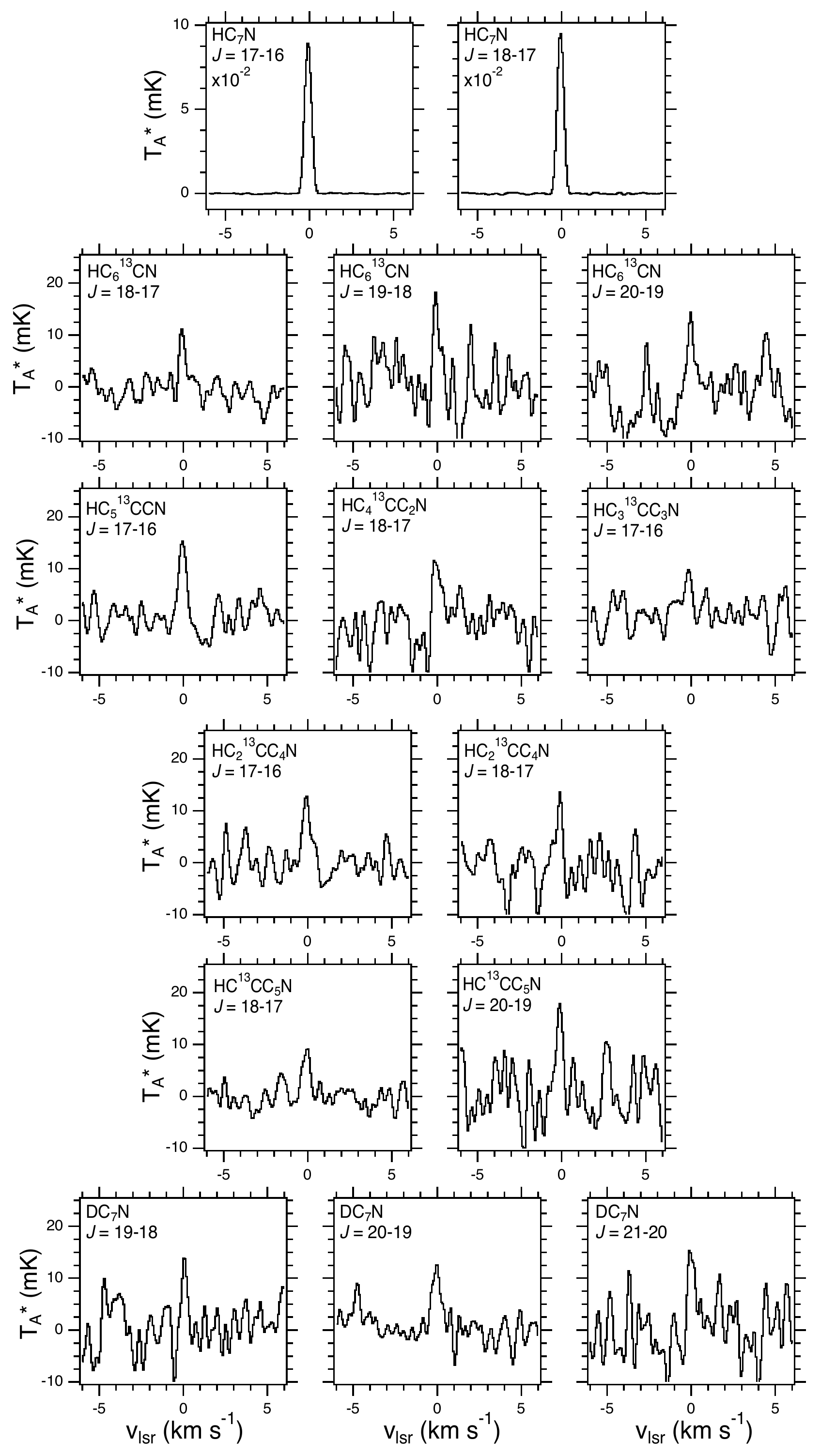}
\caption{Detected transitions of HC$_7$N isotopologues, organized by isotope-location and corresponding quantum numbers for each transition (labelled on in top left of each spectra). Velocities are given with respect to $V_{\text{lsr}}$ of the transition rest frequency, with the listed transition centred at 0 km s$^{-1}$.}
\label{fig:spectra}
\end{figure}

\begin{figure}
\centering
\includegraphics[width=0.5\textwidth]{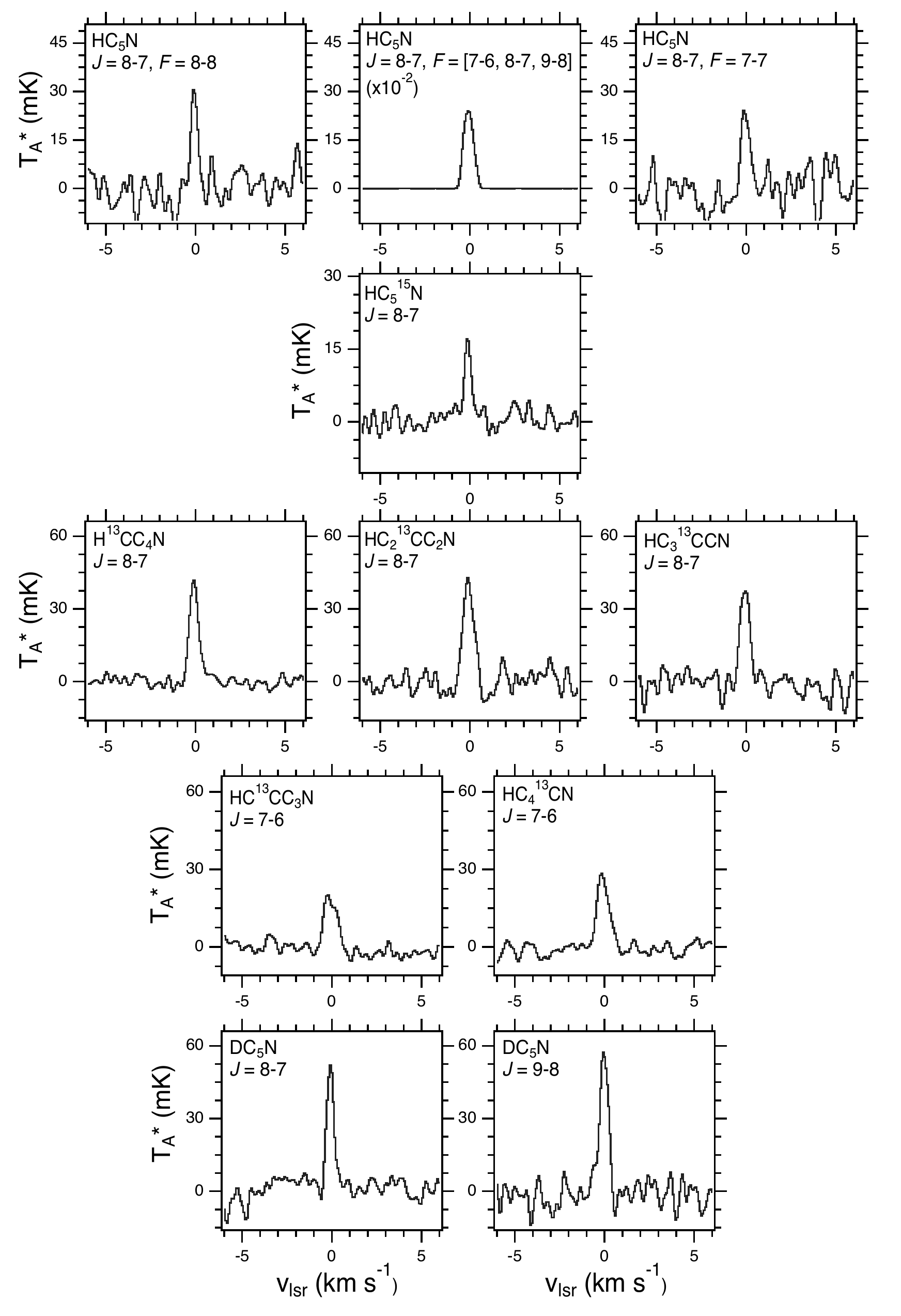}
\caption{Detected transitions of HC$_5$N isotopologues, organized by isotope-location and corresponding quantum numbers for each transition (labelled on in top left of each spectra). Velocities are given with respect to $V_{\text{lsr}}$ of the transition rest frequency, with the listed transition centred at 0 km s$^{-1}$.}
\label{fig:hc5n_spectra}
\end{figure}

We detected, for the first time, emission from DC$_7$N and six $^{13}$C-bearing isotopologues of HC$_7$N. In addition, we confirm the recent detection of HC$_5$$^{15}$N \citep{Taniguchi:2017a} with the observation of the $J$=8$\rightarrow$7 transition, which was not reported in that work. For the $^{13}$C-isotopomers, at least one $\Delta J$ transition between 17$\rightarrow$16 and 20$\rightarrow$19 was detected for six of the seven isotopomers. In addition, two transitions of DC$_5$N, two transitions of HC$_7$N, and a set of 5 hyperfine components of a single $\Delta J$ for HC$_5$N were also detected. Spectra for these species are shown in Figures \ref{fig:spectra} and \ref{fig:hc5n_spectra}. 
The lines are seen at a $v_{lsr}$\,$\sim$5.8 km s$^{-1}$, typical of molecules in this source \citep{Kaifu:2004tk}. For the one non-detected isotopomer, H$^{13}$CC$_6$N, an upper limit on the column density was derived whose value is consistent with the other detected isotopomers.

In TMC-1, the molecular emission can be well described by a single excitation temperature between $T_{ex}\sim$5-10 K  \citep{Remijan:2006tw,Loomis:2016js}. To calculate the column density, we use the formalism described in \cite{Hollis:2004uh} and given by
\begin{equation}
N_T = \frac{Q e^{E_u/T_{ex}}}{\frac{8\pi^3}{3k_B} \nu S_{ij} \mu^2} \times \frac{\frac{1}{2}  \sqrt{\frac{\pi}{\ln2}} \frac{\Delta T_A^* \Delta V}{\eta_B} }{1 -   \frac{e^{h\nu/k_B T_{ex}} -1 }{e^{h\nu/k_B T_{bg}} -1 }     } \label{eq:nt}.
\end{equation}
Here $N_T$ is the column density (cm$^{-2}$), $Q$ is the partition function (see below), $E_u$ is the upper state energy of a given transition (K), $T_{ex}$ is the excitation temperature (K), $\nu$ is the transition rest frequency (Hz), $S_{ij}$ is the intrinsic line strength, $\mu^2$ is the transition dipole moment squared (J cm$^3$), $\Delta T_A^*$ is the peak intensity (K), $\Delta V$ is the fitted FWHM linewidth (cm s$^{-1}$), $\eta_B$ is the beam efficiency ($\sim$\added{0.92} at 20 GHz for the GBT), and $T_{bg}$ is the continuum background temperature (2.73 K). Because of the narrow range of upper level energies of the observed transitions, we assume that $T_{ex} = 7$ K for all species.

The total partition function $Q$ accounts for both the rotational and vibrational contributions, as described by 
\begin{equation}
Q = Q_{\text{vib}}\times Q_{\text{rot}} \label{eq:q}.
\end{equation}
While the rotational component dominates at interstellar conditions, cyanopolyynes can have $Q$ values that can be affected by the vibrational component at even modest excitation temperatures. We utilised the calculated harmonic stretching vibrational wavenumbers ($\omega$) for HC$_7$N by \cite{Botschwina:1997}, with the assumption that the partition function will not be significantly impacted by the presence or location of $^{13}$C or D in the molecule.
The lowest three energy levels for HC$_7$N are 62, 163, and 280 cm$^{-1}$ (92, 241, and 415 K, respectively). At 7 K, the change from the vibrational contribution is negligible $\left( \Delta Q/Q \left[7\text{ K} \right]\sim 10^{-6}\right)$; this correction becomes $\gtrsim$1\% at $T_{ex}\gtrsim 20$ K. Similar behavior is seen for HC$_5$N isotopologues.

Values of $\Delta T_A^*$ and $\Delta V$ were determined by Gaussian fits to the lines (see Table \ref{tab:transitions}). For species with more than one transition, a single column density was obtained based on a least-squares fit to reproduce the integrated line intensities, with a weighting based on the SNR of the lines. The calculated column densities are summarized in Table \ref{tab:nt}.

\begin{table*}
\centering
\footnotesize
\caption{Measured column densities and upper limits for isotopomers and isotopologues discussed here, and, when relevant, the isotopic ratio    (i.e. $^{12}$C/$^{13}$C; H/D; $^{14}$N/$^{15}$N). }
\begin{tabular}{@{\hskip3pt}c@{\hskip3pt} c c c @{\hskip1pt}c@{\hskip1pt}}
\hline\hline
Species				&		$N_T$			&	$N_T/N_{\text{H}_2}$	&	Isotopic Ratio$^{\dagger}$  &  $\mathcal{R}$$^{\ddagger}$	\\
					&($\times10^{11}$ cm$^{-2}$)	&	($\times10^{-11}$)			&		\\
\hline
HC$_7$N	&	139(36)	&	139(36)	&	 -  	&   -	 \\
\\
HC$_6$$^{13}$CN	&	1.7(6)	&	1.7(6)	&	83(34)	&	85(35)   \\
HC$_5$$^{13}$CCN	&	2.6(7)	&	2.6(7)	&	52(20)	&	53(20)   \\
HC$_4$$^{13}$CC$_2$N	&	2.1(7)	&	2.1(7)	&	66(27)	&	67(28)   \\
HC$_3$$^{13}$CC$_3$N	&	1.6(4)	&	1.6(4)	&	88(32)	&	90(33)   \\
HC$_2$$^{13}$CC$_4$N	&	1.8(5)	&	1.8(5)	&	78(31)	&	79(32)   \\
HC$^{13}$CC$_5$N	&	2.0(7)	&	2.0(7)	&	71(31)	&	73(31)   \\
H$^{13}$CC$_6$N	&	$<$2.2	&	$<$2.2	&	$>$63	&	$>$64   \\
\\
DC$_7$N	&	2.5(9)	&	2.5(9)	&	56(24)	&	61(29)   \\
\hline
Weighted $^{13}$C Average Value	&	1.9(2)		&	1.9(2)				&	73(21)	& 75(21) \\
Total $^{13}$C-isotopologue$^{\star}$	&	13.3(1.8)			&	13.3(1.8)		& 9.6(2.7)\%$^{\ominus}$	& 9.4(2.7)\%$^{\ominus}$		\\
\hline
HC$_5$N	&	492(122)	&	492(122)	&	 - 	&	 -   \\
\\
HC$_5$$^{15}$N	&	1.5(4)	&	1.5(4)	&	326(109)   	&	344(114)   \\
\\
HC$_4$$^{13}$CN	&	4.6(9)	&	4.6(9)	&	107(35)   	&	108(36)   \\
HC$_3$$^{13}$CCN	&	4.4(9)	&	4.4(9)	&	111(37)   	&	113(38)   \\
HC$_2$$^{13}$CC$_2$N	&	5.3(1.2)	&	5.3(1.2)	&	93(31)   	&	94(32)   \\
HC$^{13}$CC$_3$N		&	3.4(7)	&	3.4(7)	&	144(47)   	&	146(49)   \\
H$^{13}$CC$_4$N		&	4.8(1.0)	&	4.8(1.0)	&	102(33)   	&	103(34)   \\
\\
DC$_5$N	&	5.3(1.2)	&	5.3(1.2)	&	92(30)   	&	96(32)   \\
\hline
Weighted $^{13}$C Average Value	&	4.4(4)			&	4.4(4)				&	111(30)	& 113(31)\\
Total $^{13}$C-isotopologue$^{\star}$	&	22(3)			&	22(3)			& 4.5(1.2)\%$^{\ominus}$	& 4.4(1.2)\%$^{\ominus}$	\\
\hline
\hline
\multicolumn{4}{l}{$^{\dagger}$ Column density ratio of most common isotopologue vs less common species given} \\
\multicolumn{4}{l}{~~~(e.g. $^{12}$C/$^{13}$C; H/D; $^{14}$N/$^{15}$N)}\\
\multicolumn{4}{l}{$^{\ddagger}$ Isotope ratio, including the all other singly-substituted isotopologues (see section \ref{sec:ratios})}\\
\multicolumn{4}{l}{$^{\star}$ Calculated by 7 times the weighted average of $^{13}$C values}\\
\multicolumn{4}{l}{$^{\ominus}$ Percent of total molecular density of HC$_n$N containing a single $^{13}$C substitution}\\
\end{tabular}
\label{tab:nt}
\end{table*}

To calculate the uncertainties, we considered both the measurements and analysis. The absolute flux calibration procedure for the GBT is estimated to have $\sim$20\% uncertainty. We include 1$\sigma$ uncertainties in $\Delta T_A^*$ and $\Delta V$ derived from the Gaussian fits. Due to our assumption of a single excitation temperature of $T_{ex}$= 7 K, we find that variations of previously calculated excitation temperatures (5-10 K) result in between 15-20\% uncertainty in the resulting column densities for HC$_7$N isotopologues and $\sim$5\% for HC$_5$N isotopologues. We assume that the source is significantly extended beyond the GBT beam, and thus the contributions to the uncertainty from pointing are trivial, and no beam filling correction is applied. The resulting column densities, with all uncertainties added in quadrature, are given in Table \ref{tab:nt}.

For the purposes of calculating the molecular abundances, we used the H$_2$ column density derived by \citet{Gratier:2016fj} from observations from \citet{Kaifu:2004tk} of $N_{\text{H}_2}$=10$^{22}$ cm$^{-2}$, with the caveat that the beam size of the survey performed with the Nobeyama 45m dish telescope is about twice that of the GBT, and thus a non-isotropic distribution of H$_2$, or any molecular species, may result in different column densities derived between the observations. These abundances are also given in Table \ref{tab:nt}. In addition to comparing the relative column densities of the isotopomers, an average value across all detected $^{13}$C-isotopomers is also calculated. The column density for each detected species was averaged, weighted by the error of each value, and are also tabulated in Table \ref{tab:nt}. These calculated values are compared to previous observations in Table~3.

\subsection{Treatment of Hyperfine Splitting}

For HC$_5$N, the brightest detected signal is a blend of three hyperfine components. Two additional, weaker $\Delta F$=0 hyperfine components are resolved. The central, bright feature has been shown to be slightly optically thick \citep{Gratier:2016fj}. Our calculation of the column density was therefore derived using the two, optically-thin hyperfine components, yielding a value of 5(1)$\times$10$^{13}$ cm$^{-2}$, in agreement with the previous work \citep{MacLeod:1981,Gratier:2016fj,Taniguchi:2016}. 

For all other species studied here, the hyperfine components are unresolved. As such, for the purposes of calculating column densities, the hyperfine splitting is not considered and the integrated intensity is used instead. This likely slightly overestimates the linewidths due to the blending of hyperfine components. However, this is still a reasonable assumption, as the lines are well-modeled by a single Gaussian lineshape, and thus will not significantly impact the resulting column densities, which are calculated via the integrated intensity in Equation \ref{eq:nt}.

\subsection{Calculation of Line Ratios and Total Values} \label{sec:ratios}
Relative isotopic ratios calculated for H/D, $^{12}$C/$^{13}$C, and $^{14}$N/$^{15}$N are tabulated in Table \ref{tab:nt}. It should be noted that none of the uncertainties described above for determining the column densities should cancel out in the calculation of the ratios, as the uncertainty in the absolute flux calibration comes from the time and frequency variability of the calibrator source. For HC$_7$N, where we lack a detection of H$^{13}$CC$_6$N, we calculate a total column density of all $^{13}$C-substituted isotopomers by scaling the average column density for the isotopologue by the number
of isotopomers, which is equal to the number of carbon atoms in the molecule.. 
 
As discussed in \citet{Langston:2007}, the presence of isotopologues affects the observed abundances of the main isotopic species, especially for increasingly complex molecules. As the number of atoms in a species increases, so does the probability that any given molecule will contain at least a single isotope-substituted atom. This would be most apparent in fullerenes like C$_{60}$, where a Galactic $^{12}$C/$^{13}$C ratio of $\sim$68, or 1.5\%, \citep{Milam:2005} would result in $\sim$60\% of all C$_{60}$ containing, at least one $^{13}$C substitution. Similarly, it may be important to consider the size of the molecule when comparing the total isotopologue fraction. Given the same ISM ratio and purely $^{13}$C-substitution, and no additional chemical bias, the larger cyanopolyynes would be expected to have the following total singly-substituted $^{13}$C fractional abundances: HC$_5$N (7.2\%), HC$_7$N (10\%),  HC$_9$N (12\%),  HC$_{11}$N (15\%).

\begin{figure*}
\includegraphics[width=\textwidth]{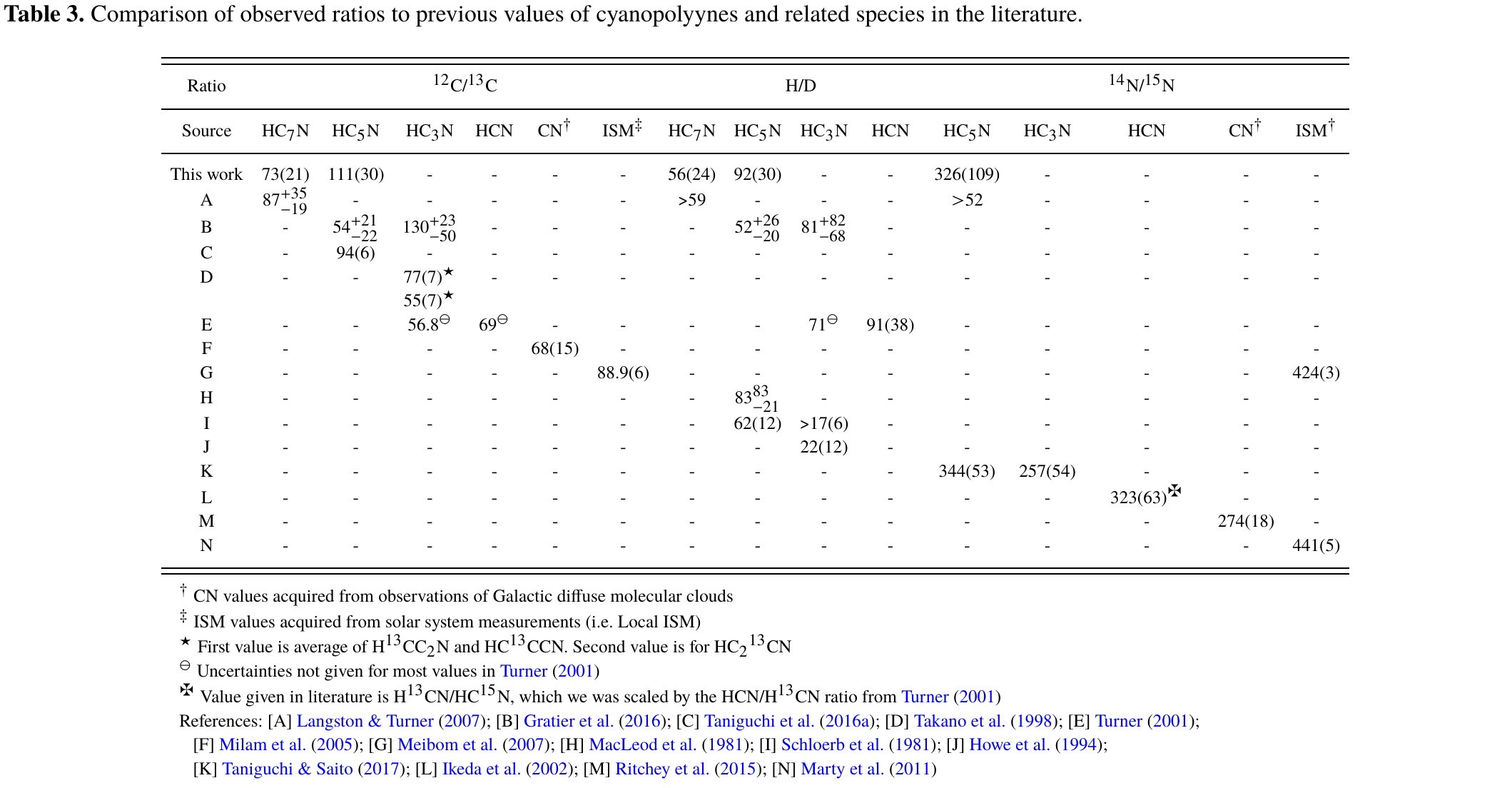}
\end{figure*}

Given the non-trival fraction of isotopologues for a given large species, as the observational capabilities improve it will likely be important to consider their abundances when calculating isotopic ratios. For the purposes of comparing to standard ISM values, a more accurate ratio for fractionation would be (taking $^{15}$N of HC$_5$N as an example)
\begin{equation}
\footnotesize
 \mathcal{R} = \frac{\text{N}_{\text{T}}(\text{HC}_5\text{N}) + \Sigma \text{N}_{\text{T}}(\text{H}^{13}\text{CC}_4\text{N isotopomers}) + \text{N}_{\text{T}}(\text{DC}_5\text{N})}{\text{N}_{\text{T}}(\text{HC}_5\text{}^{15}\text{N})}.
\end{equation}
Here, it is assumed that doubly-substituted isotopologues do not contribute significantly yet to the ratio, which may not be valid for species as large as C$_{60}$. In addition to the standard column density ratio given, this additional value is also given in Table \ref{tab:nt}, even though the relative differences ($\sim$10\%) are still well within the observational uncertainties. As chemical models increase their molecular complexity, it will become increasing important to consider the isotopologues of large molecules, as well as their precursor species. This is already shown by \citep{Majumdar:2017} on the modeling of deuterated species toward TMC-1, where the robust inclusion of deuterated chemistry was found to alter the chemical time-scales by up to a factor of 3. This effect will also be more apparent when species contain at least one atom with a higher natural percentage in non-standard isotopes, such as S and Cl and species common in silicate grain precursors (Mg, Si, Ti, Fe). While the upper limit of this effect will likely decrease the modeled abundances by no more than roughly factor of 2, it is certainly an important effect to consider in the future.

\section{Discussion} \label{sec:discussion}  

Constraining the chemistry for larger cyanopolyynes requires the consideration of both the possible formation routes, as described in \S\ref{sec:intro}, and the many precursor species, whose isotopic ratios across the literature are summarized in Table~3. The analysed isotopic ratios in this work agree reasonably well with previous observations or lower limits for the same species. The general spread in values for a given species may be attributable to the spread in excitation temperatures used (5-10 K) or inconsistent treatments of both the optical depths and hyperfine structure in the column density calculations for HC$_5$N and HC$_7$N. For HC$_3$N, the relative enhancement of $^{12}$C/$^{13}$C for HC$_2$$^{13}$CN and the agreement of the $^{14}$N/$^{15}$N ratio with Galactic measurements of CN both indicate that this species may be efficiently produced through the reaction containing CN, as described in Mechanism \ref{eq:cn} \citep{Takano:1998}. \citet{Taniguchi:2016} showed that HC$_5$N does not show this same trend. Through the comparison of the various isotopologues studied here, it is further possible to differentiate among the three mechanisms described in \S\ref{sec:intro} for both HC$_5$N and HC$_7$N.

In contrast to the $\sim$40\% enhancement of HC$_2$$^{13}$CN over HC$^{13}$CCN and H$^{13}$CC$_2$N \citep{Takano:1998}, the relative isotopic ratios of all $^{13}$C isotopomers for both HC$_7$N and HC$_5$N were all consistent with their respective average values within our uncertainties. The ratio of column densities of the HC$_{n-1}$$^{13}$CN isotopomers to the weighted average $^{13}$C-isotopolouge values can be used to more explicitly test the significance of Mechanism 1 in the formation of larger cyanopolyynes. Specifically the column densities of HC$_4$$^{13}$CN and HC$_6$$^{13}$CN agree with their respective $^{13}$C average values to $<$10\%. Because both HC$_5$N and HC$_7$N do not display the HC$_{n-1}$$^{13}$CN enhancement, this provides evidence that larger cyanopolyynes are not produced from CN to the same extent as HC$_3$N. Even though H$^{13}$CC$_6$N was not detected here, the formation route from CN should not significantly alter the $^{12}$C/$^{13}$C isotopic ratio of this molecule and so does not conflict with this conclusion. Thus, we can eliminate Mechanism \ref{eq:cn} as the dominant pathway for both HC$_5$N and HC$_7$N.

As seen Table~3, while still in agreement within the uncertainties, all isotopologues of HC$_5$N are found to be depleted (i.e. larger isotopic ratios) relative to HC$_3$N and HC$_7$N, showing that this isotopic depletion seen in \citet{Taniguchi:2016} does not continue for HC$_7$N.  While the average $^{12}$C/$^{13}$C ratio for HC$_7$N agrees very well with ratios corresponding to HCN, Galactic measurements of CN, and the two non-enhanced isotopomers of HC$_3$N, the average HC$_5$N $^{12}$C/$^{13}$C ratio is $\sim$50\% larger than any of these values. For HC$^{15}$N, the $^{14}$N/$^{15}$N ratio is much larger compared to HC$_3$N and Galactic measurements of CN, which have been shown to have enhanced $^{15}$N isotopologue abundances relative to average ISM values \citep{Roueff:2015, Ritchey:2015, Hily-Blant:2013}. Similarly, the DC$_5$N abundance is diminished by roughly the same percentage relative to HC$_7$N as the $^{13}$C-substituted isotopologues in our data. However, large uncertainties due to the SNR of our data and inconsistent treatment of the excitation temperature and hyperfine splitting across the literature results in a large spread of H/D for any given species, making this trend less certain than for the $^{12}$C/$^{13}$C and $^{14}$N/$^{15}$N ratios. Considering all of this, the mixture of  formation and destruction methods for cyanopolyynes appears to not be consistent across the molecular family.

More specifically, because the $^{12}$C/$^{13}$C ratio in HC$_5$N is not in agreement with the ratios observed for HC$_3$N and HC$_7$N, the carbon fractionation in these species does not appear to be inherited from the next-smallest cyanopolyyne, as would be predicted by Mechanism \ref{eq:hcn-2n}. Given that Mechanism \ref{eq:hcn-2n} is inefficient for HC$_5$N and HC$_7$N production, the only remaining formation pathway proposed is the reaction of hydrocarbon ions with nitrogen atoms (Mechanism \ref{eq:n}), and thus is the best prediction for the dominant production route for large cyanopolyynes. If the trends discussed here are found to be true, a dedicated investigation of the underlying chemistry of hydrocarbon ions and undetected C$_n$H isotopologues may reveal a unique $^{13}$C distribution and provide constraints on the formation of cyanopolyynes and other carbon-chain molecules.  

\section{Conclusions}
The interstellar detections of DC$_7$N, six of the seven $^{13}$C-bearing isotopologues of HC$_7$N, and HC$_5$$^{15}$N are reported toward TMC-1 with observations using the GBT. Column densities for each of the detected species and an upper limit for H$^{13}$CC$_6$N, were calculated, as well as the resulting isotopic ratios for each species. From analysis of these ratios, we find that:
\begin{itemize}
	\item There are no significant $^{12}$C/$^{13}$C variations among the isotopomers of both HC$_5$N and HC$_7$N, implying that CN is not an important precursor for their formation. 
	\item For all isotopologues studied in this work, while the values still agree within our uncertainties, HC$_5$N is found to be isotopically depleted relative to other HC$_n$N molecules and this trend does not continue onto HC$_7$N. Given also that the $^{13}$C and $^{15}$N ratios for HC$_3$N and HC$_7$N agree very well, there is evidence that cyanopolyynes are not efficiently formed from their next-smallest molecular family member, HC$_{n-2}$N 
	\item As a result, the only remaining significant formation route for HC$_5$N and HC$_7$N is the reaction of hydrocarbon ions and nitrogen atoms
\end{itemize}

\section*{Acknowledgements}
A.M.B. is a Grote Reber Fellow, and support for this work was provided by the NSF through the Grote Reber Fellowship Program administered by Associated Universities, Inc./National Radio Astronomy Observatory and the Virginia Space Grant Consortium. E. H. thanks the National Science Foundation for support of his astrochemistry program. S.V.K. acknowledges support from Basic Research Program P-7 of the Presidium of the Russian Academy of Sciences. A.M.B. thanks C.N. Shingledecker for helpful discussions on chemical models for cyanopolyynes. B.A.M. thanks K. L. Lee for helpful discussions regarding vibrational energy levels. The National Radio Astronomy Observatory is a facility of the National Science Foundation operated under cooperative agreement by Associated Universities, Inc. The Green Bank Observatory is a facility of the National Science Foundation operated under cooperative agreement by Associated Universities, Inc. The authors thank the anonymous referee for comments that improved the quality of this manuscript.



\begin{thebibliography}{}
\makeatletter
\relax
\def\mn@urlcharsother{\let\do\@makeother \do\$\do\&\do\#\do\^\do\_\do\%\do\~}
\def\mn@doi{\begingroup\mn@urlcharsother \@ifnextchar [ {\mn@doi@}
  {\mn@doi@[]}}
\def\mn@doi@[#1]#2{\def\@tempa{#1}\ifx\@tempa\@empty \href
  {http://dx.doi.org/#2} {doi:#2}\else \href {http://dx.doi.org/#2} {#1}\fi
  \endgroup}
\def\mn@eprint#1#2{\mn@eprint@#1:#2::\@nil}
\def\mn@eprint@arXiv#1{\href {http://arxiv.org/abs/#1} {{\tt arXiv:#1}}}
\def\mn@eprint@dblp#1{\href {http://dblp.uni-trier.de/rec/bibtex/#1.xml}
  {dblp:#1}}
\def\mn@eprint@#1:#2:#3:#4\@nil{\def\@tempa {#1}\def\@tempb {#2}\def\@tempc
  {#3}\ifx \@tempc \@empty \let \@tempc \@tempb \let \@tempb \@tempa \fi \ifx
  \@tempb \@empty \def\@tempb {arXiv}\fi \@ifundefined
  {mn@eprint@\@tempb}{\@tempb:\@tempc}{\expandafter \expandafter \csname
  mn@eprint@\@tempb\endcsname \expandafter{\@tempc}}}

\bibitem[\protect\citeauthoryear{{Araki}, {Takano}, {Sakai}, {Yamamoto},
  {Oyama}, {Kuze}  \& {Tsukiyama}}{{Araki} et~al.}{2016}]{Araki:2016}
{Araki} M.,  {Takano} S.,  {Sakai} N.,  et al.,  2016, \mn@doi [\apj] {10.3847/1538-4357/833/2/291}, \href
  {http://adsabs.harvard.edu/abs/2016ApJ...833..291A} {833, 291}

\bibitem[\protect\citeauthoryear{{Bell}, {Avery}, {MacLeod}  \&
  {Matthews}}{{Bell} et~al.}{1992}]{Bell:1992}
{Bell} M.~B.,  {Avery} L.~W.,  {MacLeod} J.~M., et al.,  1992,
  \mn@doi [\apj] {10.1086/172017}, \href
  {http://cdsads.u-strasbg.fr/abs/1992ApJ...400..551B} {400, 551}

\bibitem[\protect\citeauthoryear{{Bell}, {Feldman}, {Travers}, {McCarthy},
  {Gottlieb}  \& {Thaddeus}}{{Bell} et~al.}{1997}]{Bell:1997vj}
{Bell} M.~B.,  {Feldman} P.~A.,  {Travers} M.~J.,  et al.,  1997, \mn@doi [\apjl] {10.1086/310732},
  \href {http://adsabs.harvard.edu/abs/1997ApJ...483L..61B} {483, L61}

\bibitem[\protect\citeauthoryear{{Benson} \& {Myers}}{{Benson} \&
  {Myers}}{1989}]{Benson:1989}
{Benson} P.~J. \&   {Myers} P.~C.,  1989, \mn@doi [\apjs] {10.1086/191365}, \href
  {http://adsabs.harvard.edu/abs/1989ApJS...71...89B} {71, 89}

\bibitem[\protect\citeauthoryear{{Bizzocchi}, {Degli Esposti}  \&
  {Botschwina}}{{Bizzocchi} et~al.}{2004}]{Bizzocchi:2004}
{Bizzocchi} L.,  {Degli Esposti} C., \&    {Botschwina} P.,  2004, \mn@doi [Journal
  of Molecular Spectroscopy] {10.1016/j.jms.2004.02.019}, \href
  {http://adsabs.harvard.edu/abs/2004JMoSp.225..145B} {225, 145}

\bibitem[\protect\citeauthoryear{{Botschwina}, {Horn}, {Markey}  \&
  {Oswald}}{{Botschwina} et~al.}{1997}]{Botschwina:1997}
{Botschwina} P.,  {Horn} M.,  {Markey} K., et al.,  1997, \mn@doi
  [Molecular Physics] {10.1080/00268979709482108}, \href
  {http://adsabs.harvard.edu/abs/1997MolPh..92..381B} {92, 381}

\bibitem[\protect\citeauthoryear{{Broten}, {Oka}, {Avery}, {MacLeod}  \&
  {Kroto}}{{Broten} et~al.}{1978}]{Broten:1978}
{Broten} N.~W.,  {Oka} T.,  {Avery} L.~W.,  et al.,
  1978, \mn@doi [\apjl] {10.1086/182739}, \href
  {http://cdsads.u-strasbg.fr/abs/1978ApJ...223L.105B} {223, L105}

\bibitem[\protect\citeauthoryear{{Bujarrabal}, {Guelin}, {Morris}  \&
  {Thaddeus}}{{Bujarrabal} et~al.}{1981}]{Bujarrabal:1981}
{Bujarrabal} V.,  {Guelin} M.,  {Morris} M., et al.,  1981, \aap,
  \href {http://adsabs.harvard.edu/abs/1981A%26A....99..239B} {99, 239}

\bibitem[\protect\citeauthoryear{{Cernicharo}, {Heras}, {Tielens}, {Pardo},
  {Herpin}, {Gu{\'e}lin}  \& {Waters}}{{Cernicharo}
  et~al.}{2001}]{Cernicharo:2001}
{Cernicharo} J.,  {Heras} A.~M.,  {Tielens} A.~G.~G.~M.,  et al.,  2001, \mn@doi [\apjl]
  {10.1086/318871}, \href {http://adsabs.harvard.edu/abs/2001ApJ...546L.123C}
  {546, L123}

\bibitem[\protect\citeauthoryear{{Fukuzawa}, {Osamura}  \&
  {Schaefer}}{{Fukuzawa} et~al.}{1998}]{Fukuzawa:1998}
{Fukuzawa} K.,  {Osamura} Y., \&   {Schaefer} III H.~F.,  1998, \mn@doi [\apj]
  {10.1086/306168}, \href {http://adsabs.harvard.edu/abs/1998ApJ...505..278F}
  {505, 278}

\bibitem[\protect\citeauthoryear{{Gratier}, {Majumdar}, {Ohishi}, {Roueff},
  {Loison}, {Hickson}  \& {Wakelam}}{{Gratier} et~al.}{2016}]{Gratier:2016fj}
{Gratier} P.,  {Majumdar} L.,  {Ohishi} M.,  et al.,  2016, \mn@doi [\apjs]
  {10.3847/0067-0049/225/2/25}, \href
  {http://adsabs.harvard.edu/abs/2016ApJS..225...25G} {225, 25}

\bibitem[\protect\citeauthoryear{{Guzman-Ramirez}, {Zijlstra},
  {N{\'{\i}}chuim{\'{\i}}n}, {Gesicki}, {Lagadec}, {Millar}  \&
  {Woods}}{{Guzman-Ramirez} et~al.}{2011}]{Guzman-Ramirez:2011}
{Guzman-Ramirez} L.,  {Zijlstra} A.~A.,  {N{\'{\i}}chuim{\'{\i}}n} R.,  et al.,  2011, \mn@doi
  [\mnras] {10.1111/j.1365-2966.2011.18502.x}, \href
  {http://adsabs.harvard.edu/abs/2011MNRAS.414.1667G} {414, 1667}

\bibitem[\protect\citeauthoryear{{Hasegawa}, {Herbst}  \& {Leung}}{{Hasegawa}
  et~al.}{1992}]{Hasegawa:1992}
{Hasegawa} T.~I.,  {Herbst} E.,  \&  {Leung} C.~M.,  1992, \mn@doi [\apjs]
  {10.1086/191713}, \href {http://adsabs.harvard.edu/abs/1992ApJS...82..167H}
  {82, 167}

\bibitem[\protect\citeauthoryear{{Herbst}}{{Herbst}}{1983}]{Herbst:1983}
{Herbst} E.,  1983, \mn@doi [\apjs] {10.1086/190882}, \href
  {http://adsabs.harvard.edu/abs/1983ApJS...53...41H} {53, 41}

\bibitem[\protect\citeauthoryear{{Herbst} \& {Leung}}{{Herbst} \&
  {Leung}}{1990}]{Herbst&Leung:1990}
{Herbst} E. \&   {Leung} C.~M.,  1990, \aap, \href
  {http://adsabs.harvard.edu/abs/1990A%26A...233..177H} {233, 177}

\bibitem[\protect\citeauthoryear{Herbst \& Millar}{Herbst \&
  Millar}{2008}]{Herbst&Millar:2008}
Herbst E. \&   Millar T.~J.,  2008, The Chemistry of Cold Interstellar Cloud Cores
  in Low Temperatures and Cold Molecules.
World Scientific

\bibitem[\protect\citeauthoryear{{Herbst}, {Adams}  \& {Smith}}{{Herbst}
  et~al.}{1984}]{Herbst:1984}
{Herbst} E.,  {Adams} N.~G., \&   {Smith} D.,  1984, \mn@doi [\apj]
  {10.1086/162538}, \href {http://adsabs.harvard.edu/abs/1984ApJ...285..618H}
  {285, 618}

\bibitem[\protect\citeauthoryear{{Hily-Blant}, {Bonal}, {Faure}  \&
  {Quirico}}{{Hily-Blant} et~al.}{2013}]{Hily-Blant:2013}
{Hily-Blant} P.,  {Bonal} L.,  {Faure} A., et al.,  2013, \mn@doi
  [\icarus] {10.1016/j.icarus.2012.12.015}, \href
  {http://adsabs.harvard.edu/abs/2013Icar..223..582H} {223, 582}

\bibitem[\protect\citeauthoryear{Hollis, Jewell, Lovas  \& Remijan}{Hollis
  et~al.}{2004}]{Hollis:2004uh}
Hollis J.~M.,  Jewell P.~R.,  Lovas F.~J.,  et al.,  2004, 
  \mn@doi [\apj] {10.1086/424927}, \href
  {http://adsabs.harvard.edu/abs/2004ApJ...613L..45H} {613, L45}

\bibitem[\protect\citeauthoryear{{Howe}, {Millar}, {Schilke}  \&
  {Walmsley}}{{Howe} et~al.}{1994}]{Howe:1994}
{Howe} D.~A.,  {Millar} T.~J.,  {Schilke} P., et al.,  1994,
  \mn@doi [\mnras] {10.1093/mnras/267.1.59}, \href
  {http://adsabs.harvard.edu/abs/1994MNRAS.267...59H} {267, 59}

\bibitem[\protect\citeauthoryear{{Huntress}}{{Huntress}}{1977}]{Huntress:1977}
{Huntress} Jr. W.~T.,  1977, \mn@doi [\apjs] {10.1086/190439}, \href
  {http://adsabs.harvard.edu/abs/1977ApJS...33..495H} {33, 495}

\bibitem[\protect\citeauthoryear{{Ikeda}, {Hirota}  \& {Yamamoto}}{{Ikeda}
  et~al.}{2002}]{Ikeda:2002}
{Ikeda} M.,  {Hirota} T., \&  {Yamamoto} S.,  2002, \mn@doi [\apj]
  {10.1086/341287}, \href {http://adsabs.harvard.edu/abs/2002ApJ...575..250I}
  {575, 250}

\bibitem[\protect\citeauthoryear{{Irvine}, {Hoglund}, {Friberg}, {Askne}  \&
  {Ellder}}{{Irvine} et~al.}{1981}]{Irvine:1981}
{Irvine} W.~M.,  {Hoglund} B.,  {Friberg} P.,  et al.,
  1981, \mn@doi [\apjl] {10.1086/183637}, \href
  {http://adsabs.harvard.edu/abs/1981ApJ...248L.113I} {248, L113}

\bibitem[\protect\citeauthoryear{{Kaifu} et~al.,}{{Kaifu}
  et~al.}{2004}]{Kaifu:2004tk}
Kaifu, N., Ohishi, M., Kawaguchi, K., et al., 2004,
\mn@doi [\pasj] {10.1093/pasj/56.1.69}, \href
  {http://adsabs.harvard.edu/abs/2004PASJ...56...69K} {56, 69}

\bibitem[\protect\citeauthoryear{{Kirkpatrick} et~al.,}{{Kirkpatrick}
  et~al.}{2017}]{Kirkpatrick:2017}
Kirkpatrick, A., Alberts, S., Pope, A.,  et~al., 2017, \mn@doi [\pasj]{10.3847/1538-4357/aa911d} \href
  {http://http://adsabs.harvard.edu/abs/2017ApJ...849..111K}{849, 111} 
  
\bibitem[\protect\citeauthoryear{{Knight}, {Freeman}, {McEwan}, {Smith}  \&
  {Adams}}{{Knight} et~al.}{1986}]{Knight:1986}
{Knight} J.~S.,  {Freeman} C.~G.,  {McEwan} M.~J.,  et al.,  1986, \mn@doi [\mnras] {10.1093/mnras/219.1.89}, \href
  {http://adsabs.harvard.edu/abs/1986MNRAS.219...89K} {219, 89}

\bibitem[\protect\citeauthoryear{{Langston} \& {Turner}}{{Langston} \&
  {Turner}}{2007}]{Langston:2007}
{Langston} G. \&  {Turner} B.,  2007, \mn@doi [\apj] {10.1086/511332}, \href
  {http://adsabs.harvard.edu/abs/2007ApJ...658..455L} {658, 455}

\bibitem[\protect\citeauthoryear{{Loomis} et~al.,}{{Loomis}
  et~al.}{2016}]{Loomis:2016js}
Loomis, R.~A., Shingledecker, C.~N., Langston, G., et al., 2016, \mn@doi [\mnras] {10.1093/mnras/stw2302}, \href
  {http://adsabs.harvard.edu/abs/2016MNRAS.463.4175L} {463, 4175}

\bibitem[\protect\citeauthoryear{{MacLeod}, {Avery}  \& {Broten}}{{MacLeod}
  et~al.}{1981}]{MacLeod:1981}
{MacLeod} J.~M.,  {Avery} L.~W., \&  {Broten} N.~W.,  1981, \mn@doi [\apjl]
  {10.1086/183687}, \href {http://adsabs.harvard.edu/abs/1981ApJ...251L..33M}
  {251, L33}

\bibitem[\protect\citeauthoryear{{Majumdar} et~al.,}{{Majumdar}
  et~al.}{2017}]{Majumdar:2017}
Majumdar, L., Gratier, P., Ruaud, M.,  et~al., 2017, \mn@doi [\mnras] {10.1093/mnras/stw3360}, \href
  {http://adsabs.harvard.edu/abs/2017MNRAS.466.4470M} {466, 4470}

\bibitem[\protect\citeauthoryear{{Marty}, {Chaussidon}, {Wiens}, {Jurewicz}  \&
  {Burnett}}{{Marty} et~al.}{2011}]{Marty:2011}
{Marty} B.,  {Chaussidon} M.,  {Wiens} R.~C.,  et al.,  2011, \mn@doi [Science] {10.1126/science.1204656}, \href
  {http://adsabs.harvard.edu/abs/2011Sci...332.1533M} {332, 1533}

\bibitem[\protect\citeauthoryear{{Mauersberger}, {Henkel}  \&
  {Sage}}{{Mauersberger} et~al.}{1990}]{Mauersberger:1990}
{Mauersberger} R.,  {Henkel} C.,  \& {Sage} L.~J.,  1990, \aap, \href
  {http://adsabs.harvard.edu/abs/1990A%26A...236...63M} {236, 63}

\bibitem[\protect\citeauthoryear{{McCarthy}, {Travers}, {Kov{\'a}cs},
  {Gottlieb}  \& {Thaddeus}}{{McCarthy} et~al.}{1997}]{McCarthy:1997uv}
{McCarthy} M.~C.,  {Travers} M.~J.,  {Kov{\'a}cs} A.,  et al.,  1997, \mn@doi [\apjs] {10.1086/313050}, \href
  {http://adsabs.harvard.edu/abs/1997ApJS..113..105M} {113, 105}

\bibitem[\protect\citeauthoryear{{McCarthy}, {Levine}, {Apponi}  \&
  {Thaddeus}}{{McCarthy} et~al.}{2000}]{McCarthy:2000eo}
{McCarthy} M.~C.,  {Levine} E.~S.,  {Apponi} A.~J.,   et al.,  2000,
  \mn@doi [JMoSp] {10.1006/jmsp.2000.8149}, \href
  {http://adsabs.harvard.edu/abs/2000JMoSp.203...75M} {203, 75}

\bibitem[\protect\citeauthoryear{{McCarthy}, {Gottlieb}, {Gupta}  \&
  {Thaddeus}}{{McCarthy} et~al.}{2006}]{McCarthy:2006tk}
{McCarthy} M.~C.,  {Gottlieb} C.~A.,  {Gupta} H.,  et al.,  2006,
  \mn@doi [\apjl] {10.1086/510238}, \href
  {http://adsabs.harvard.edu/abs/2006ApJ...652L.141M} {652, L141}

\bibitem[\protect\citeauthoryear{{McGuire}, {Burkhardt}, {Shingledecker},
  {Kalenskii}, {Herbst}, {Remijan}  \& {McCarthy}}{{McGuire}
  et~al.}{2017}]{McGuire:2017}
{McGuire} B.~A.,  {Burkhardt} A.~M.,  {Shingledecker} C.~N.,  et al.,  2017, \mn@doi
  [\apjl] {10.3847/2041-8213/aa7ca3}, \href
  {http://adsabs.harvard.edu/abs/2017ApJ...843L..28M} {843, L28}

\bibitem[\protect\citeauthoryear{{Meibom}, {Krot}, {Robert}, {Mostefaoui},
  {Russell}, {Petaev}  \& {Gounelle}}{{Meibom} et~al.}{2007}]{Meibom:2007}
{Meibom} A.,  {Krot} A.~N.,  {Robert} F.,  et al.,  2007, \mn@doi [\apjl] {10.1086/512052},
  \href {http://adsabs.harvard.edu/abs/2007ApJ...656L..33M} {656, L33}

\bibitem[\protect\citeauthoryear{{Milam}, {Savage}, {Brewster}, {Ziurys}  \&
  {Wyckoff}}{{Milam} et~al.}{2005}]{Milam:2005}
{Milam} S.~N.,  {Savage} C.,  {Brewster} M.~A.,  et al.,  2005, \mn@doi [\apj] {10.1086/497123}, \href
  {http://adsabs.harvard.edu/abs/2005ApJ...634.1126M} {634, 1126}

\bibitem[\protect\citeauthoryear{{Ohishi} \& {Kaifu}}{{Ohishi} \&
  {Kaifu}}{1998}]{Ohishi&Kaifu:1998}
{Ohishi} M. \& {Kaifu} N.,  1998, \mn@doi [Faraday Discuss.]
  {10.1039/a801058g}, \href {http://adsabs.harvard.edu/abs/1998FaDi..109..205O}
  {109, 205}

\bibitem[\protect\citeauthoryear{{Remijan}, {Wyrowski}, {Friedel}, {Meier}  \&
  {Snyder}}{{Remijan} et~al.}{2005}]{Remijan:2005}
{Remijan} A.~J.,  {Wyrowski} F.,  {Friedel} D.~N.,  et al.,  2005, \mn@doi [\apj] {10.1086/429750}, \href
  {http://adsabs.harvard.edu/abs/2005ApJ...626..233R} {626, 233}

\bibitem[\protect\citeauthoryear{{Remijan}, {Hollis}, {Snyder}, {Jewell}  \&
  {Lovas}}{{Remijan} et~al.}{2006}]{Remijan:2006tw}
{Remijan} A.~J.,  {Hollis} J.~M.,  {Snyder} L.~E.,  et al.,  2006, \mn@doi [\apjl] {10.1086/504918}, \href
  {http://adsabs.harvard.edu/abs/2006ApJ...643L..37R} {643, L37}

\bibitem[\protect\citeauthoryear{{Ritchey}, {Federman}  \& {Lambert}}{{Ritchey}
  et~al.}{2015}]{Ritchey:2015}
{Ritchey} A.~M.,  {Federman} S.~R., \&  {Lambert} D.~L.,  2015, \mn@doi [\apjl]
  {10.1088/2041-8205/804/1/L3}, \href
  {http://adsabs.harvard.edu/abs/2015ApJ...804L...3R} {804, L3}

\bibitem[\protect\citeauthoryear{{Roueff}, {Loison}  \& {Hickson}}{{Roueff}
  et~al.}{2015}]{Roueff:2015}
{Roueff} E.,  {Loison} J.~C., \&  {Hickson} K.~M.,  2015, \mn@doi [\aap]
  {10.1051/0004-6361/201425113}, \href
  {http://adsabs.harvard.edu/abs/2015A\%26A...576A..99R} {576, A99}

\bibitem[\protect\citeauthoryear{{Ruaud}, {Wakelam}  \& {Hersant}}{{Ruaud}
  et~al.}{2016}]{Ruaud:2016}
{Ruaud} M.,  {Wakelam} V.,  \& {Hersant} F.,  2016, \mn@doi [\mnras]
  {10.1093/mnras/stw887}, \href
  {http://adsabs.harvard.edu/abs/2016MNRAS.459.3756R} {459, 3756}

\bibitem[\protect\citeauthoryear{{Schiff} \& {Bohme}}{{Schiff} \&
  {Bohme}}{1979}]{Schiff&Bohme:1979}
{Schiff} H.~I. \& {Bohme} D.~K.,  1979, \mn@doi [\apj] {10.1086/157334}, \href
  {http://adsabs.harvard.edu/abs/1979ApJ...232..740S} {232, 740}

\bibitem[\protect\citeauthoryear{{Schloerb}, {Snell}, {Langer}  \&
  {Young}}{{Schloerb} et~al.}{1981}]{Schloerb:1981}
{Schloerb} F.~P.,  {Snell} R.~L.,  {Langer} W.~D.,   et al.,  1981,
  \mn@doi [\apjl] {10.1086/183688}, \href
  {http://adsabs.harvard.edu/abs/1981ApJ...251L..37S} {251, L37}

\bibitem[\protect\citeauthoryear{{Snyder}, {Hollis}, {Jewell}, {Lovas}  \&
  {Remijan}}{{Snyder} et~al.}{2006}]{Snyder:2006}
{Snyder} L.~E.,  {Hollis} J.~M.,  {Jewell} P.~R., et al.,  2006, \mn@doi [\apj] {10.1086/505323}, \href
  {http://adsabs.harvard.edu/abs/2006ApJ...647..412S} {647, 412}

\bibitem[\protect\citeauthoryear{{Takano} et~al.,}{{Takano}
  et~al.}{1998}]{Takano:1998}
Takano, S., Masuda, A., Hirahara, Y.,  et~al., 1998, \aap, \href
  {http://adsabs.harvard.edu/abs/1998A%26A...329.1156T} {329, 1156}

\bibitem[\protect\citeauthoryear{{Taniguchi} \& {Saito}}{{Taniguchi} \&
  {Saito}}{2017}]{Taniguchi:2017a}
{Taniguchi} K. \& {Saito} M.,  2017, \mn@doi [\pasj]
  {10.1093/pasj/psx065}, \href
  {http://adsabs.harvard.edu/abs/2017PASJ...69L...7T} {69, L7}

\bibitem[\protect\citeauthoryear{{Taniguchi}, {Ozeki}, {Saito}, {Sakai},
  {Nakamura}, {Kameno}, {Takano}  \& {Yamamoto}}{{Taniguchi}
  et~al.}{2016a}]{Taniguchi:2016}
{Taniguchi} K.,  {Ozeki} H.,  {Saito} M.,  et al.,  2016a, \mn@doi [\apj]
  {10.3847/0004-637X/817/2/147}, \href
  {http://adsabs.harvard.edu/abs/2016ApJ...817..147T} {817, 147}

\bibitem[\protect\citeauthoryear{{Taniguchi}, {Saito}  \& {Ozeki}}{{Taniguchi}
  et~al.}{2016b}]{Taniguchi:2016b}
{Taniguchi} K., {Saito} M., \&  {Ozeki} H.,  2016b, \mn@doi [\apj]
  {10.3847/0004-637X/830/2/106}, \href
  {http://adsabs.harvard.edu/abs/2016ApJ...830..106T} {830, 106}

\bibitem[\protect\citeauthoryear{{Taniguchi}, {Ozeki}  \& {Saito}}{{Taniguchi}
  et~al.}{2017}]{Taniguchi:2017b}
{Taniguchi} K.,  {Ozeki} H.,  \& {Saito} M.,  2017, \mn@doi [\apj]
  {10.3847/1538-4357/aa82ba}, \href
  {http://adsabs.harvard.edu/abs/2017ApJ...846...46T} {846, 46}

\bibitem[\protect\citeauthoryear{{Turner}}{{Turner}}{2001}]{Turner:2001}
{Turner} B.~E.,  2001, \mn@doi [\apjs] {10.1086/322536}, \href
  {http://adsabs.harvard.edu/abs/2001ApJS..136..579T} {136, 579}

\bibitem[\protect\citeauthoryear{{Ulich} \& {Haas}}{{Ulich} \&
  {Haas}}{1976}]{Ulich&Haas:1976}
{Ulich} B.~L. \& {Haas} R.~W.,  1976, \mn@doi [\apjs] {10.1086/190361}, \href
  {http://adsabs.harvard.edu/abs/1976ApJS...30..247U} {30, 247}

\bibitem[\protect\citeauthoryear{{Watson}, {Anicich}  \& {Huntress}}{{Watson}
  et~al.}{1976}]{Watson:1976}
{Watson} W.~D.,  {Anicich} V.~G.,  \& {Huntress} Jr. W.~T.,  1976, \mn@doi
  [\apjl] {10.1086/182115}, \href
  {http://adsabs.harvard.edu/abs/1976ApJ...205L.165W} {205, L165}

\bibitem[\protect\citeauthoryear{{Winnewisser} \& {Walmsley}}{{Winnewisser} \&
  {Walmsley}}{1979}]{Winnewisser&Walmsley:1979}
{Winnewisser} G. \&  {Walmsley} C.~M.,  1979, \mn@doi [\apss]
  {10.1007/BF00643491}, \href
  {http://adsabs.harvard.edu/abs/1979Ap%26SS..65...83W} {65, 83}

\makeatother
\end{thebibliography}
\end{document}